\documentclass[10pt,conference]{IEEEtran}
\IEEEoverridecommandlockouts %

\usepackage[utf8]{inputenc}
\usepackage[T1]{fontenc}
\usepackage{amsmath,amssymb}
\usepackage{graphicx}
\usepackage{booktabs}
\usepackage{xcolor}
\usepackage{microtype}
\usepackage{tikz}
\usetikzlibrary{arrows.meta,positioning}
\usepackage[hidelinks]{hyperref}

\begin{document}
\bstctlcite{IEEEexample:BSTcontrol}

\title{Benchmark Factory: Agentic Benchmark Creation via\\Dataset Curation and Label-Free Quality Anchors}

\author{\IEEEauthorblockN{Shervin Shahidi\textsuperscript{*}}
\IEEEauthorblockA{\textit{Meta Platforms}\\
shervin@meta.com
}
\and
\IEEEauthorblockN{James Poon\textsuperscript{*}}
\IEEEauthorblockA{\textit{Meta Platforms}\\
jypoon@meta.com
}
\and
\IEEEauthorblockN{Hang Xie}
\IEEEauthorblockA{\textit{Meta Platforms}\\
hangxie@meta.com
}
\and
\IEEEauthorblockN{Kenneth O. Xu}
\IEEEauthorblockA{\textit{Meta Platforms}\\
kennethoxu@meta.com
}
\and
\IEEEauthorblockN{Robin H{\aa}kanson}
\IEEEauthorblockA{\textit{Meta Platforms}\\
matshakanson@meta.com
}
\and
\IEEEauthorblockN{Vikram Nair}
\IEEEauthorblockA{\textit{Meta Platforms}\\
vikramnair@meta.com
}
\and
\IEEEauthorblockN{Jens Pillgram-Larsen}
\IEEEauthorblockA{\textit{Meta Platforms}\\
jenspl@meta.com
}
\and
\IEEEauthorblockN{Nachiappan Nagappan}
\IEEEauthorblockA{\textit{Meta Platforms}\\
nnachi@meta.com
}
\thanks{\textsuperscript{*}These authors contributed equally to this work}}

\maketitle

\begin{abstract}
Teams deploying coding agents need benchmarks built from their own code, but curating and validating them by hand is resource intensive, and an unvalidated benchmark can mislead silently. We present Benchmark Factory, an autonomous meta-agent loop with human checkpoints: an LLM agent writes sourcing queries over landed production code changes, runs models of known relative strength on each candidate, diagnoses why a candidate fails, and repairs it by editing its queries, prompts, and configuration until it clears quality anchors that need no per-task human labels.
We used Benchmark Factory to build benchmarks for five developer cohorts at our organization. Engineering raters scored a 5\% sample 2.3 for task quality and 2.6 for grader verdicts on a 1-to-3 scale. On configurations never used to build them, the benchmarks ranked Claude Opus~4.7 at high reasoning effort above low effort (60.6\% vs.\ 54.4\% of tasks passed, $p=0.002$) and GPT-5.5 above GPT-5.4 (65.8\% vs.\ 60.5\%, $p=0.004$); 9 of the 10 per-cohort comparisons pointed the same way. A deliberately trivial benchmark failed the anchors; five broken graders failed them, even under a relaxed rule fixed in advance that the real grader passes; and with model labels shuffled so that no true ordering existed, random redraws of a candidate passed in at most 1.4\% of simulations. The anchors are necessary, not sufficient, conditions for a valid benchmark, designed to focus human review on what the loop cannot repair.
\end{abstract}

\section{Introduction}
Software teams increasingly delegate code changes to AI coding agents, and they need to measure those agents on their own code: their applications, operating systems, and internal frameworks. Public suites such as SWE-bench~\cite{jimenez2024swebench} do not cover this work, and curated benchmarks with careful outcome checks, such as PerfBench~\cite{garg2026perfbench}, rely on expert verification of each task, which is hard to repeat for every team and codebase. Generating benchmarks automatically from a team's history of code changes fills the gap but creates a new problem: a benchmark that is too easy, too hard, or too noisy silently distorts every comparison made on it, and checking it by hand is resource intensive enough that it is often skipped.

Benchmark Factory makes benchmark construction an agentic task. A meta-agent, an LLM agent that builds benchmarks for other agents, writes the evaluation configuration and the query that sources candidate tasks from landed code changes, launches runs of a ladder of models of known relative strength, and reads per-task results and trajectories. It accepts a candidate benchmark only if the outcome agrees with relations that hold without human labels (quality anchors): a clearly stronger model should pass more tasks than a weaker one, at pass rates where models can be told apart. Otherwise it diagnoses the failure and repairs the candidate, preferring changes to the data over changes to how it is scored (Sec.~\ref{sec:method}).

A benchmark is accepted when the very models that define the expected ordering line up on it, and the loop repairs it until they do, so passing could reflect those models' quirks rather than coding capability. Passing the anchors is therefore only a necessary condition, and we test what it is worth with two hypotheses on evidence the loop never saw. \textbf{H1}: accepted benchmarks rank configurations they were not built against in the expected direction. \textbf{H2}: benchmarks and graders known to be broken fail the anchors.

\textbf{Contributions.} (1) Benchmark Factory, an autonomous meta-agent loop gated by configurable label-free anchors (Sec.~\ref{sec:method}); (2) benchmarks built for five developer cohorts (Sec.~\ref{sec:build}); (3) evidence that accepted benchmarks behave as intended: they rank unseen configurations as expected, and an expert audit rates tasks 2.3 and verdicts 2.6 of 3, with under-specified tasks the main failure (Sec.~\ref{sec:right}); stability across runs and graders is checked on companion benchmarks built by the same loop; (4) evidence that the anchors reject broken benchmarks (Sec.~\ref{sec:h2}); and (5) lessons for building such loops (Sec.~\ref{sec:disc}).

The loop has been used to build further benchmarks beyond the five studied here; we evaluate the five that cover a diverse set of developer activities and populations, and for which we have done a thorough analysis, including out-of-sample runs and expert review.

\section{Benchmark Factory}
\label{sec:method}
Benchmark Factory uses our internal agent-evaluation platform, which runs any coding agent and grades its output with tests or an LLM.

\textbf{Tasks and benchmarks.} A task is a landed (merged) production code change: the agent receives the change's title and summary and works at its parent commit. Because tasks come from private repositories, public models are unlikely to have seen their landed patches in training. When the change ships with tests, the task is SWE-bench-style~\cite{jimenez2024swebench} and the tests grade the agent's patch; otherwise an LLM grader compares the patch with the landed one. A benchmark is a sourcing query over such changes, together with the task prompt, the grader, and its pass threshold.

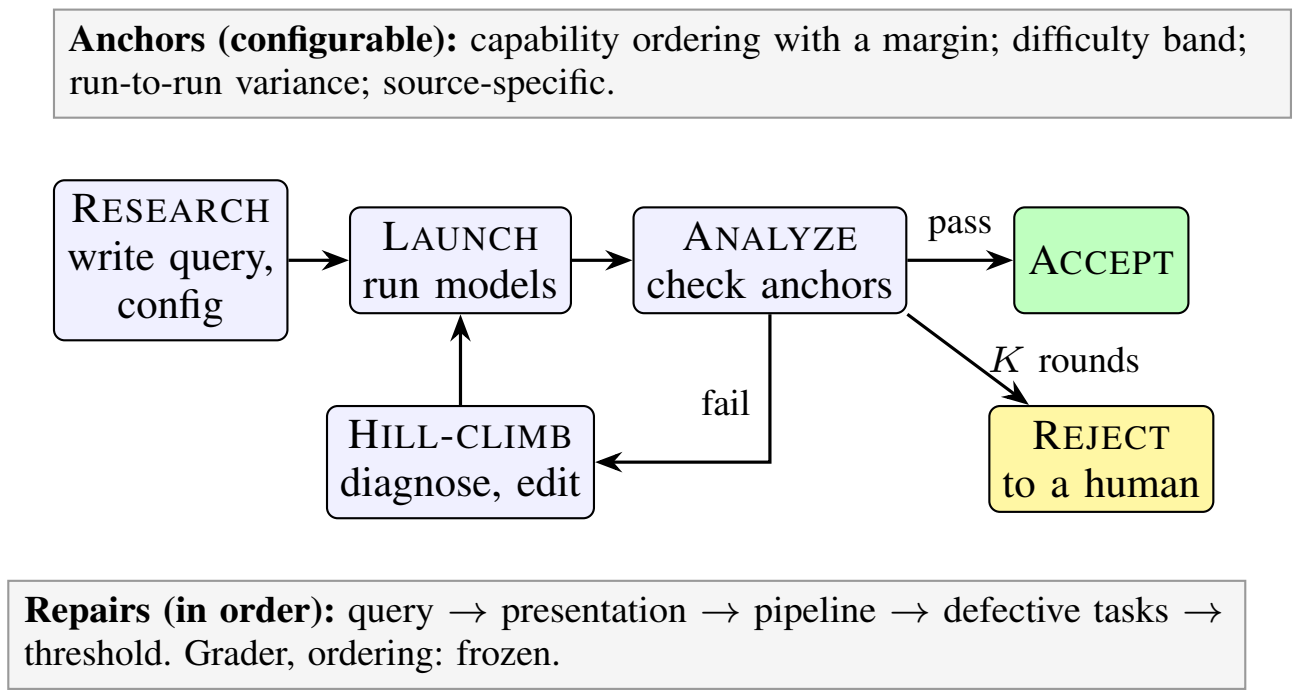
\begin{figure}[t]
\centering
\begin{tikzpicture}[font=\footnotesize, node distance=5mm and 4mm,
  st/.style={draw, rounded corners=2pt, minimum height=7mm, inner sep=2.5pt, align=center, fill=blue!6},
  ann/.style={draw=black!40, fill=black!4, inner sep=3pt, align=left, font=\scriptsize, text width=0.9\columnwidth},
  ar/.style={-{Stealth[length=1.8mm]}, semithick}]
\node[ann] (chk) {\textbf{Anchors (configurable):} capability ordering with a margin; difficulty band; run-to-run variance; source-specific.};
\node[st, below=4mm of chk.south west, anchor=north west] (res) {\textsc{Research}\\write query,\\config};
\node[st, right=of res] (lau) {\textsc{Launch}\\run models};
\node[st, right=of lau] (ana) {\textsc{Analyze}\\check anchors};
\node[st, right=7mm of ana, fill=green!25] (acc) {\textsc{Accept}};
\node[st, below=6mm of lau] (hc) {\textsc{Hill-climb}\\diagnose, edit};
\node[st, below=6mm of acc, fill=yellow!45] (esc) {\textsc{Reject}\\to a human};
\draw[ar] (res) -- (lau); \draw[ar] (lau) -- (ana); \draw[ar] (ana) -- node[above, font=\scriptsize]{pass} (acc);
\draw[ar] (ana) |- node[pos=0.3, left, font=\scriptsize]{fail} (hc); \draw[ar] (hc) -- (lau);
\draw[ar] (ana.south east) -- node[right, font=\scriptsize, pos=0.5]{$K$ rounds} (esc);
\node[ann, below=4mm of hc.south, anchor=north, xshift=11mm] {\textbf{Repairs (in order):} query $\rightarrow$ presentation $\rightarrow$ pipeline $\rightarrow$ defective tasks $\rightarrow$ threshold. Grader, ordering: frozen.};
\end{tikzpicture}
\caption{The Benchmark Factory loop for one candidate. Every phase is executed by the LLM meta-agent; repairs act on the hill-climb subset, and the holdout never drives a repair.}
\label{fig:loop}
\vspace{-10pt}
\end{figure}

\textbf{The loop} (Fig.~\ref{fig:loop}). A driver invokes the meta-agent in a fresh context each iteration; it performs one step and keeps its state in a file, so long runs do not degrade (the ``Ralph'' loop~\cite{huntley2025ralph}). Each candidate moves through four phases. In \textsc{Research} the agent explores production data (landed changes, data tables, documentation) and writes and validates the sourcing query and evaluation configuration. In \textsc{Launch} it starts one run per model under test. In \textsc{Analyze} it separates infrastructure failures from genuine ones, checks the anchors, and accepts the candidate, sends it to \textsc{Hill-climb}, or rejects it after $K$ rounds and hands it to a human, who decides whether the domain does not separate the models, the pipeline is faulty, or the expected ordering is wrong.

\textbf{Quality anchors.} The anchors are an extensible set of a-priori checks on the pass rates $r_m$ of the models under test; they are label-free in that they need no per-task human labels, although the tiers and checkpoints are human inputs. (i) \emph{Capability ordering}: the models are grouped into tiers by release order and public results, and $r_a - r_b \ge \delta$ for every model $a$ in a higher tier than model $b$. Models in one tier are left unordered, since adjacent models are often indistinguishable at realistic sample sizes. The margin $\delta$ relaxes a significance test, which benchmarks of about 100 tasks rarely pass; paired tests are reported but do not gate. Because the margin is a policy choice rather than a noise bound, a candidate for which it is clearly too strict or too loose (small sample, close tiers, volatile runs) goes to human review, which may adjust it. (ii) \emph{Difficulty band}: every $r_m$ lies in an informative range, to avoid saturation of scores. (iii) \emph{Low run-to-run variance}, when repeated runs are available. A data source can add its own anchors; a code-review judge benchmark, for instance, added floors on precision and recall in predicting whether a reviewed change would land.

\textbf{Holdout.} \textsc{Research} splits each candidate, by hashing task identifiers, into a larger hill-climb subset, on which anchors are checked and repairs made, and a holdout that runs with every round. The meta-agent sees the holdout's results, but its membership is frozen by a hash re-checked every iteration, and by rule it never gates acceptance or drives a repair; it is reported at acceptance as a regression check. A repair that re-sources the whole pool may draw a new one.

\textbf{Hill-climb.} The meta-agent diagnoses which anchor failed and why, from task-level results and trajectories, and edits the candidate accordingly. Repairs are ordered so that those least able to tailor the benchmark to the models come first: the agent rewrites the sourcing query (e.g., toward larger or more recent changes), then changes how tasks are presented, fixes the pipeline, and inspects tasks every model passes or fails, dropping one only for a defect in the task and encoding it as a sourcing rule; the pass threshold moves last. Infrastructure failures go straight to pipeline fixes. The grader and the expected ordering are not repairs: they change only on evidence from outside the candidate, such as an execution-based check, and are then frozen. Otherwise the loop could satisfy its anchors by tuning the measurement.

\section{Building Benchmarks for Five Cohorts}
\label{sec:build}
\textbf{Setup.} We used Benchmark Factory to build benchmarks for five developer cohorts in our organization, Cohorts 1--5 (C1--C5), each a group of teams working in one product area, spanning several programming languages and both monorepo and multi-repo settings. No fail-to-pass tests are recorded for them, so an LLM grader (GPT-5.5) scores the agent's patch against the landed patch from 0 to 100. The models under test were Claude Opus~4.7 in the top tier, Opus~4.6 and Sonnet~4.6 in the middle tier, and Opus~4.5 in the bottom tier; C5, built a month later, used Opus~4.8 $\succ$ Opus~4.7 $\succ$ \{Opus~4.6, Sonnet~4.6\}. The anchors used $\delta=0.05$ and a band of $[0.10, 0.85]$ (no variance anchor, since rounds were not repeated), with a 20\% holdout and rejection after five rounds; a second consecutive threshold drop needed human approval, and a human reviewed every acceptance.

\begin{table}[t]
\centering
\caption{The five cohort benchmarks as accepted. HC/HO: hill-climb and holdout tasks; $t$: pass threshold; Rnd: repair rounds to acceptance; Changes: number of changes of each type. Pass rates: top tier / Opus~4.6, Sonnet~4.6 / bottom tier.}
\label{tab:cohorts}
\setlength{\tabcolsep}{3pt}
\scriptsize
\begin{tabular}{@{}lcccl ll@{}}
\toprule
Cohort & HC/HO & $t$ & Rnd & Changes & Hill-climb rates & Holdout rates \\
\midrule
C1 & 96/24 & 80 & 6$^{a}$ & S1 T3 G3 F2 & .64 / .48, .57 / .41 & n/r \\
C2 & 96/24 & 60 & 1 & S1 & .67 / .59, .59 / .35 & .86 / .68, .68 / .35 \\
C3 & 96/24 & 75 & 4 & S3 P1 T1 & .82 / .66, .72 / .59 & .79 / .70, .79 / .59 \\
C4 & 144/36 & 75 & 3 & S3$^{b}$ F1 T1 & .66 / .52, .52 / .38 & .62 / .48, .43 / .38$^{d}$ \\
C5 & 52/-- & 75 & 0$^{c}$ & -- & .81 / .75 / .58, .60 & -- \\
\bottomrule
\end{tabular}\\[2pt]
\raggedright\scriptsize S sourcing query, P prompt, T threshold, G grader or expected ordering, F infrastructure fix (S3 = three sourcing changes); n/r: not recorded. $^{a}$Development cohort, run past the five-round limit under human supervision while the rules were revised. $^{b}$Infrastructure-motivated (repository reachability, pipeline errors). $^{c}$A new candidate after an earlier one was rejected; 52 paired tasks, no holdout; tiers Opus~4.8 / Opus~4.7 / Opus~4.6, Sonnet~4.6. $^{d}$On a 21-task re-run.
\vspace{-10pt}
\end{table}

\textbf{What the loop did.} Across the five cohorts the loop produced five accepted benchmarks and rejected one candidate (C5); humans intervened in C1, C4, and C5. Four of the five accepted benchmarks needed repair (Table~\ref{tab:cohorts}): 8 sourcing changes, 5 threshold changes, 3 infrastructure fixes, and 1 prompt change, plus three grader or ordering changes in C1. In C2, at round 0 Opus~4.7 and Sonnet~4.6 passed 70\% and 68\% of tasks ($p=0.66$), too close for the margin; the meta-agent rewrote the sourcing query toward larger changes (3--6 files, at least 50 lines), which separated them (67\% vs.\ 59\%), replacing 81 of the 96 tasks by a stated criterion rather than re-scoring the same ones. Threshold changes served only the band. In C3 and C4 the ordering already held at the original threshold of 60, but the top model sat above the band's ceiling; raising the threshold to 75, which lowers every model's pass rate, fixed only that. C3 confirmed it on a fresh round; C4 re-thresholded its round's stored grader scores and confirmed on a fresh holdout run. Mean grader scores, where recorded (C1, C2, C5), keep the accepted ordering (C5's top step ties).

C2 and C3 kept one holdout throughout, while C1 re-drew it once and C4 three times, when re-sourcing changed the population. On these small holdouts (21--24 graded tasks) the top tier stays above the bottom tier in every recorded case, and C2 keeps the full ordering; in C3 Sonnet~4.6 ties the top model, and in C4 one pair falls 0.002 short of the margin. C5 shows a rejection. No worker could initially check out all of its repositories, so the first candidate was narrowed to five projects; it kept the bottom-tier model above a middle-tier one across two rounds and on its holdout (combined $p=0.049$, descriptive). The inversion persisted, and the candidate was rejected as an operational decision.
Once the infrastructure covered the full population, a fresh attempt on the new dataset, not a further round of the rejected one, passed without repair; its tiers were set from the then-current models (Opus~4.8 added, Opus~4.5 dropped), so the inverted pair was not re-tested.

\textbf{Effort.} A round runs four models on about 120 tasks (some 480 agent runs) and, depending on the level of concurrency, can finish within a few hours. C2 was accepted after two rounds (about 960 agent runs) and C3 after five (about 2{,}400). The exact cost depends on the agents under test, which run the tasks; the meta-agent itself takes only a few iterations per round.

\section{Do Accepted Benchmarks Behave as Intended?}
\label{sec:right}

\subsection{Expert review}
\label{sec:human}
We drew a 5\% sample of the roughly 600 tasks in the cohorts' pools, six per cohort at random (30 tasks), and re-ran it with the newest model (Opus~4.8) and a bottom-tier one (Opus~4.5; Opus~4.6 for C5). Three engineering experts from our organization (one rater per task) rated each task's quality as poor (1), acceptable (2), or good (3), its difficulty, and each grader verdict as wrong (1), arguable (2), or correct (3), seeing the task, both patches, and the grader's reasoning. Task quality averaged 2.3: 15 tasks were good, 9 acceptable, and 6 poor. Verdict quality averaged 2.6: of 58 verdicts (two runs produced no gradable output), 40 were correct (69\%; 95\% CI 53--83\%, bootstrap over tasks), 13 arguable, and 5 wrong (9\%).

The wrong verdicts share one cause. No PASS verdict was rated wrong; all five wrong verdicts were FAILs on tasks whose description omitted details the landed patch contained, such as enum values or item priorities, so the grader penalized the agent for details the task never specified. Most arguable verdicts follow the same pattern. Four of the six poor tasks are unclear or under-specified, and two over-specify the change. The anchors cannot catch such problems when they lower every model's pass rate alike; they call for a sourcing filter that removes or rewrites under-specified tasks (future work).

\subsection{H1: transfer to unseen configurations}
\label{sec:h1}
After acceptance we froze each benchmark definition (query, prompt, grader, threshold) and drew a new pool of 42 to 150 tasks per cohort, independently of the construction sets (2 tasks in C2 and 58 in C4 overlap by chance). On each cohort's final accepted benchmark we varied two factors that played no part in construction: an unseen model family, GPT-5.4 versus GPT-5.5, both run through a different agent harness, and the reasoning effort of Opus~4.7, which construction ran only at the agent's default. The two primary contrasts, GPT-5.4 versus GPT-5.5 and low versus high effort, use paired exact McNemar tests (one-sided; directions fixed before the runs: newer model better, pass rate non-decreasing with effort) and cohort-stratified bootstrap 95\% intervals (10{,}000 resamples). Runs that produced no output count as failures; pooled over the five cohorts, they were 10\% of GPT-5.4 and 7\% of GPT-5.5 runs and 12--13\% at every effort setting.

\begin{figure}[t]
\centering
\includegraphics[width=\columnwidth]{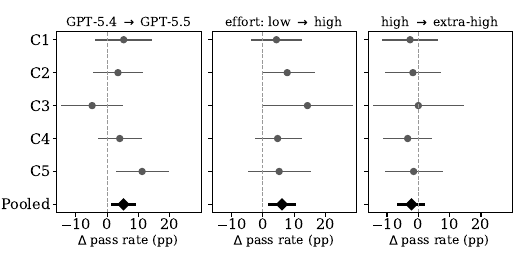}
\caption{Out-of-sample paired pass-rate deltas with 95\% bootstrap intervals, per cohort and pooled (stratified by cohort; $n=551$--$552$). None of these configurations was used to build or accept the benchmarks.}
\label{fig:oos}
\vspace{-10pt}
\end{figure}

\textbf{Results.} GPT-5.5 passes 65.8\% of tasks and GPT-5.4 60.5\%, a gain of $+5.3$ points (95\% CI $[+1.6, +8.9]$, $p=0.004$, $n=552$), ahead in four of five cohorts (Fig.~\ref{fig:oos}): benchmarks accepted against Claude models recover the release order of another vendor's models, run through a different harness. At high effort Opus~4.7 passes 60.6\% of tasks and at low effort 54.4\%, a gain of $+6.2$ points (95\% CI $[+2.2, +10.2]$, $p=0.002$, $n=551$) that is positive in all five cohorts and also appears in the mean grader score ($+2.0$ points on the 0--100 scale, $p=0.006$). Both survive a Bonferroni correction, the removal of the development cohort C1 ($+5.2$, $p=0.009$; $+6.6$, $p=0.003$), and the removal of tasks that overlap construction ($+5.1$, $p=0.009$; $+7.1$, $p<0.001$). Leaving out any single cohort keeps both gains positive (GPT $+3.4$ to $+6.1$, the smallest, without C5, at $p=0.07$; effort $+5.5$ to $+6.7$). Counting only tasks where both GPT models produced a patch, the GPT gain is $+3.9$ points ($p=0.02$, $n=465$).

GPT-5.5 trails only in C3 ($-4.8$ points), a small ($n=42$), near-saturated slice where four tasks separate the models, split 3 to 1 (two-sided $p=0.63$), and the mean grader score slightly favors GPT-5.5; per-cohort samples are underpowered, so H1 rests on the pooled contrasts.

\textbf{Extra-high effort} gives no gain over high ($-2.2$ points, not significant) despite 20\% more turns and 31\% more output tokens. On the 430 tasks graded under both settings the gap is under one point; 8 of the 12 net tasks are runs that ended without a patch, which happens at a similar rate under every setting (12--13\%). We studied effort further on two companion benchmarks built by the same loop for another organization (105 and 150 tasks, every setting run three times, three graders, a rubric with a scope check): higher effort raised the overall score but produced more patches larger than the reference and lowered the scope check, a penalty that is right when the extra changes are unwarranted and unfair when they are useful work the reference lacks.

\subsection{Stability and dependence on the grader}
\label{sec:stab}
In the companion benchmarks, graded by two runs of GPT-5.5 and one of Claude Opus~4.8, up to one verdict in six flipped between identical runs, yet repetition explained almost none of the score variation, which came mostly from the task (65--73\%) and the grader (11--24\%). The same grader re-run agreed with itself (Scott's $\pi=0.81$); the cross-family grader ranked the agents identically (all paired contrasts $p<0.001$) but scored about 11.5 points higher. In these benchmarks the grader's family moved the level, not the order; thresholds must be re-calibrated when the grader changes, which is why it is frozen with the benchmark.
In this study the grader is GPT-5.5, so self-preference could favor GPT-5.5 in the model-family contrast, but not in the effort contrast, where both arms are the same Claude model.

\section{H2: Do the Anchors Reject Broken Benchmarks?}
\label{sec:h2}
\begin{table}[t]
\centering
\caption{Negative controls ($n=52$; 51 for the self-report and no-reference graders; 57 for trivial changes). Rates: top tier / Opus~4.6, Sonnet~4.6 / bottom tier. Pairs: cross-tier pairs ordered correctly by mean grader score. Grader rows re-score the October re-run of C2 under the replay rule.}
\label{tab:neg}
\setlength{\tabcolsep}{2.4pt}
\scriptsize
\begin{tabular}{@{}l l c l@{}}
\toprule
Setup (what is broken) & Rates & Pairs & Rejected by \\
\midrule
Real grader (replay reference)$^{a}$ & .73 / .65, .71 / .64 & 5/5 & -- \\
\midrule
Interchangeable models (labels & \multicolumn{2}{l}{accepts $\le$0.22\% / $\le$1.4\%$^{b}$} & ordering \\
\quad shuffled within task) & \multicolumn{2}{l}{October re-run: 0.07\% / 0.51\%} & \\
Trivial changes (1 file, $\le$5 lines) & .95 / .91, .93 / .91 & 4/5 & band, ordering \\
Shuffled references & .00 / .00, .00 / .00 & 0/5 & band, ordering \\
Self-report grader (no code) & .96 / .94, .96 / .90 & 5/5 & band \\
No-reference grader & 1.0 / .98, 1.0 / .90 & 5/5 & band \\
Length-only score & .65 / .54, .58 / .54 & 3/5 & ordering \\
Lenient threshold (10/100) & 1.0 / .98, .98 / .98 & 5/5 & band \\
\bottomrule
\end{tabular}\\[2pt]
\raggedright\scriptsize $^{a}$Margin met on 3 of 5 pairs (all 5 ordered on score), so the unchanged anchors would reject it too. $^{b}$Accepted in one attempt / within five, worst cohort, $\delta=0.05$; 96-task draws (52 on the re-run).
\vspace{-10pt}
\end{table}

Each control in Table~\ref{tab:neg} is broken in a known way, and the anchors reject all of them. On the 52-task re-run they also reject the real grader, a low-sample case the gate sends to human review (Sec.~\ref{sec:method}); every broken grader still fails a relaxed replay rule, fixed before re-scoring, that the real grader passes. A final probe tests the grader directly.

\textbf{Interchangeable models.} We treat the four Opus~4.7 effort runs of each cohort as four models and shuffle their results within each task, which keeps task difficulty and noise but removes any true ordering, then check the anchors on 96-task draws ($20{,}000$ simulations; each attempt is a fresh draw, up to five as in the loop's round budget, without repairs). Under this null the shuffled ladder passes in at most 0.22\% of single attempts and 1.4\% within five, so random redraws rarely manufacture an ordering; the loop's adaptive repairs are checked by H1 instead. The margin trades false accepts for false rejects: at $\delta=0.03$ the five-attempt rate rises to 3.3--4.9\%, while $\delta=0.05$ can fail a good benchmark on a small sample.

\textbf{Trivial changes.} A candidate sourced with a deliberately poor query (single-file changes of 1--5 lines, $n=57$) has pass rates of 0.91--0.95, with at most three tasks separating any pair of models: it carries almost no capability signal and fails both band and ordering.

\textbf{Broken graders.} Re-scoring needs the agents' patches, which construction did not store, so in October 2026 we re-ran the C2 models on 56 tasks from C2's new pool with Opus~4.8 at the top (52 tasks have results for all four). On these new tasks, with a newer top model and harness, the real grader orders all five cross-tier pairs correctly on mean score but meets the pass-rate margin on only three (the other gaps are 0.019): only 3 and 7 of the 52 tasks separate these pairs, too few to resolve a 5-point margin. The replay rule relaxes the margin: a setup is rejected if it fails the band, or fails the ordering anchor and ranks at most three of five pairs correctly by mean grader score. Shuffled references score everything zero. A self-report grader (the agent's account of its work, never the code) and a no-reference grader (patch versus task description only) keep the ordering but saturate (0.90--1.00), failing the band. A length-only score barely tracks capability (Kendall $\tau=0.18$ versus $0.91$ for the real grader), and lowering the pass threshold to 10/100 lets nearly every patch pass (98--100\%), failing the band.

\textbf{Degenerate submissions.} On 142 tasks from C1--C3 the grader passes the landed patch 100\% of the time; a comment-only no-op, another task's patch, and size-matched padding 0\%; and a patch with only the first half of the landed change's files 8--28\%, between the extremes.

\textbf{What the anchors do not guarantee.} Only the band rejects the self-report and no-reference graders, so a free threshold could let one through (the self-report grader passes at exactly $t=94$); the loop's thresholds all lay between 60 and 80, and only the threshold's place as the last repair and human review guard against this.

\section{Discussion}
\label{sec:disc}

\textbf{Use case: evaluating a new tool.} The anchors also let a team stand up a benchmark for the question at hand. To test two Claude Code skills, we built a 40-task benchmark in one round, with a grader that also reads the trajectory; it passed the anchors (Opus~4.8 76\%, 4.7 62\%, 4.6 and Sonnet~4.6 54\%) within an hour. With Opus~4.8, a cosmetic skill that makes the agent narrate in Early Modern English moved the pass rate from 76\% to 66\%, not statistically significant ($n=38$ paired tasks; bootstrap 95\% CI $-24$ to $0$ points; exact McNemar $p=0.22$), while a sabotage skill that makes no edits and then claims success dropped it to 0\% ($p<10^{-4}$), the grader naming the false claim every time. Benchmark and grader were fixed before any run; this is an illustrative demo.

\textbf{Lessons for building agent loops.} In the development cohort, C1, Opus~4.5 scored 0\% because its model identifier was invalid. The band anchor caught this, but the meta-agent read it as low capability and lowered the threshold from 80 to 60; the ladder compressed (three rates within 0.012), and the resulting acceptance, made before the gate checked for compression, was retracted. Four rules followed: check run health (turns, tool calls, tokens) before reading a score; order repairs by their power to fit the measurement, threshold last; freeze the grader and expected ordering unless external evidence justifies a change; and keep a human checkpoint at every acceptance and rejection.

\textbf{Capability-alignment bias.} Benchmark$^2$~\cite{benchmarksq2026} treats items that a weaker model solves and a stronger one fails as flaws in existing benchmarks; when building one, the opposite risk matters: a required ordering could remove such tasks even when they reflect genuine strengths. Anchors constrain aggregate pass rates, so such tasks stay; repairs change the population by stated rules, removing tasks only for defects; and close models share a tier, so when C1 showed Sonnet~4.6 at least matching Opus~4.6 under three graders and an execution-based check, their tier was merged rather than the benchmark re-sourced. The out-of-sample axes (Sec.~\ref{sec:h1}) check that sourcing did not tilt the benchmark toward newer models' strengths.

\textbf{Limitations and future work.} All cohorts come from one organization, and no hand-curated suite exists for these codebases; the expert-reviewed sample, our substitute, is small and single-rated, so no inter-rater agreement can be estimated; raters saw model names and the grader's reasoning, which could bias verdict ratings toward the grader. The capability ordering is an assumption that the out-of-sample axes mitigate but do not remove, and the expert audit covers accepted benchmarks only, which still contained under-specified tasks. A pass rate from one grader does not capture every dimension of agent behavior~\cite{lima2026catalogue}. The model-family transfer is scored by a grader of the same family, though the companion benchmarks suggest the family shifts the level rather than the order. Holdouts are small, were re-drawn when the population changed, and were not recorded for C1's final round. We do not claim that every benchmark the loop produces is valid. Future work includes sourcing harder, higher-quality tasks, filtering or rewriting under-specified ones, and grading that separates useful extra work from scope drift.

\section{Related Work and Conclusion}
\textbf{Related work.} REAP curates coding-agent tasks from production developer--agent interactions and landed changes, checking fail-to-pass tests for relevance and stability~\cite{jha2026reap}; we target developer cohorts, grade with an LLM against the landed patch, and gate acceptance on model-order and difficulty anchors. Agentic construction with verification appears for embodied~\cite{jiang2026embodiedbenchforge} and medical~\cite{fu2026medbenchagent} benchmarks, with repair in~\cite{jiang2026embodiedbenchforge}, and AutoBencher searches for salient, difficult datasets~\cite{li2024autobencher}. BenchBuilder~\cite{li2024benchbuilder} validates a finished benchmark by separability and agreement with human preferences; Farchi et al.~\cite{farchi2024benchmarks} generate code benchmarks and their judge from synthetic artifacts. SWE-bench Verified screened every instance by hand~\cite{jimenez2024swebench, openai2024swebenchverified}, whereas our audit samples tasks and verdicts. LLM graders have known biases~\cite{ye2024justice}.

\textbf{Conclusion.} An autonomous meta-agent can build coding benchmarks from production changes and check them with label-free quality anchors. Accepted benchmarks transferred to configurations they were not built against and broken ones failed (graders even under a rule relaxed for the small sample), so verification rests on testable predictions rather than trust in the generator.

\textbf{Data availability.} The benchmarks contain internal code and cannot be released; the procedure, anchor parameters, and analyses are described here.

\bibliographystyle{IEEEtran}
\bibliography{IEEEabrv,paper}

\begin{thebibliography}{10}
\providecommand{\url}[1]{#1}
\csname url@samestyle\endcsname
\providecommand{\newblock}{\relax}
\providecommand{\bibinfo}[2]{#2}
\providecommand{\BIBentrySTDinterwordspacing}{\spaceskip=0pt\relax}
\providecommand{\BIBentryALTinterwordstretchfactor}{4}
\providecommand{\BIBentryALTinterwordspacing}{\spaceskip=\fontdimen2\font plus
\BIBentryALTinterwordstretchfactor\fontdimen3\font minus
  \fontdimen4\font\relax}
\providecommand{\BIBforeignlanguage}[2]{{%
\expandafter\ifx\csname l@#1\endcsname\relax
\typeout{** WARNING: IEEEtran.bst: No hyphenation pattern has been}%
\typeout{** loaded for the language `#1'. Using the pattern for}%
\typeout{** the default language instead.}%
\else
\language=\csname l@#1\endcsname
\fi
#2}}
\providecommand{\BIBdecl}{\relax}
\BIBdecl

\bibitem{jimenez2024swebench}
C.~E. Jimenez, J.~Yang, A.~Wettig \emph{et~al.}, ``{SWE}-bench: Can language
  models resolve real-world {GitHub} issues?'' in \emph{Proc. Int. Conf. Learn.
  Represent. (ICLR)}, 2024, pp. 54\,107--54\,157.

\bibitem{garg2026perfbench}
S.~Garg, R.~Z. Moghaddam, and N.~Sundaresan, ``{PerfBench}: Can agents resolve
  real-world performance bugs?'' in \emph{Proc. Int. Workshop Agentic Eng.
  (AGENT)}.\hskip 1em plus 0.5em minus 0.4em\relax ACM, 2026, pp. 165--172.

\bibitem{huntley2025ralph}
G.~Huntley, ``{Ralph Wiggum} as a `software engineer',''
  \url{https://ghuntley.com/ralph/}, 2025, blog post.

\bibitem{benchmarksq2026}
Q.~Qian, C.~Huang, J.~Xu \emph{et~al.}, ``Benchmark$^2$: Systematic evaluation
  of {LLM} benchmarks,'' \emph{arXiv preprint arXiv:2601.03986}, 2026.

\bibitem{lima2026catalogue}
I.~Lima, V.~Linhares, A.~M. Gomes, and P.~H. Maia, ``A catalogue of evaluation
  metrics for {LLM}-based multi-agent frameworks in software engineering,'' in
  \emph{Proc. Int. Workshop Agentic Eng. (AGENT)}.\hskip 1em plus 0.5em minus
  0.4em\relax ACM, 2026, pp. 120--124.

\bibitem{jha2026reap}
S.~Jha, M.~Paltenghi, C.~Maddila \emph{et~al.}, ``{REAP}: Automatic curation of
  coding agent benchmarks from interactive production usage,'' in \emph{Proc.\
  IEEE/ACM Int.\ Conf.\ Autom.\ Softw.\ Eng.\ (ASE)}, 2026, pp. 4304--4314.

\bibitem{jiang2026embodiedbenchforge}
B.~Jiang, F.~Zhang, L.~Wang \emph{et~al.}, ``{Embodied-BenchForge}: A
  closed-loop agentic workflow for embodied benchmark construction,''
  \emph{arXiv preprint arXiv:2609.13082}, 2026.

\bibitem{fu2026medbenchagent}
Y.~Fu, J.~Wang, G.~Yang \emph{et~al.}, ``{MedBenchAgent}: Towards systematic
  automation of medical {VLM} benchmark construction,'' \emph{arXiv preprint
  arXiv:2610.11312}, 2026.

\bibitem{li2024autobencher}
X.~L. Li, F.~Kaiyom, E.~Liu, Y.~Mai, P.~Liang, and T.~Hashimoto,
  ``{AutoBencher}: Towards declarative benchmark construction,'' in \emph{Proc.
  Int. Conf. Learn. Represent. (ICLR)}, 2025, pp. 94\,357--94\,386.

\bibitem{li2024benchbuilder}
T.~Li \emph{et~al.}, ``From crowdsourced data to high-quality benchmarks:
  {Arena-Hard} and {BenchBuilder} pipeline,'' in \emph{Proc. Int. Conf. Mach.
  Learn. (ICML)}, ser. Proceedings of Machine Learning Research, vol.
  267.\hskip 1em plus 0.5em minus 0.4em\relax PMLR, 2025, pp. 34\,209--34\,231.

\bibitem{farchi2024benchmarks}
E.~Farchi, S.~Froimovich, R.~Katan, and O.~Raz, ``Automatic generation of
  benchmarks and reliable {LLM} judgment for code tasks,'' \emph{arXiv preprint
  arXiv:2410.21071}, 2024.

\bibitem{openai2024swebenchverified}
{OpenAI}, ``Introducing {SWE}-bench {Verified},''
  \url{https://openai.com/index/introducing-swe-bench-verified/}, 2024,
  {OpenAI} blog post.

\bibitem{ye2024justice}
J.~Ye \emph{et~al.}, ``Justice or prejudice? {Q}uantifying biases in
  {LLM}-as-a-judge,'' in \emph{Proc. Int. Conf. Learn. Represent. (ICLR)},
  2025, pp. 102\,351--102\,390.

\end{thebibliography}

\end{document}